\input{aipcheck}

\documentclass[
    ,final            
  ]
  {aipproc}

\layoutstyle{6x9}

\makeatletter
\def\slash#1{{\mathpalette\c@ncel{#1}}} 
\makeatother

\newcommand\beq{\begin{eqnarray}}
\newcommand\eeq{\end{eqnarray}}

\newcommand\la{\langle}
\newcommand\ra{\rangle}

\begin{document}

\title{Impact of the triple-gluon correlation functions on the single
spin asymmetries in pp collisions}


\author{Yuji Koike}{
  address={Department of Physics, Niigata University,
Ikarashi, Niigata 950-2181, Japan}
}

\author{Shinsuke Yoshida}{
  address={Graduate School of Science and Technology, Niigata University,
Ikarashi, Niigata 950-2181, Japan}
}

\begin{abstract}
We calculate the single-spin-dependent cross section formula
for the $D$-meson production and the direct-photon production
in the $pp$ collision induced by the twist-3 triple-gluon correlation functions
in the transversely polarized nucleon.
We also present a model calculation for the asymmetries
in comparison with the preliminary data given by RHIC, 
showing the impact of the correlation functions
on the asymmetries.  
\end{abstract}

\maketitle


\section{1. Introduction}

Understanding the origin of the large single spin asymmetries (SSAs) observed in various
high-energy semi-inclusive processes have been a big challenge during
the past decades. The SSA can be generated as a consequence of the
multiparton correlations inside the hadrons in the collinear
factorization approach which is valid when the transverse momentum of final
state hadron can be regarded as hard. Recently the mesurement of SSA for heavy meson
production by the PHENIX collaboration\,\cite{Liu}
have motivated theoretical works for multigluon correlation
inside the transversly polarized proton which is represented by
the triple-gluon correlation functions\,\cite{KQVY08,BKTY10} because
heavy quarks fragmenting into final state meson are mainly produced by the
gluon fusion mechanism.

In this work, we study the contribution of the triple-gluon correlation
functions to SSA for the $D$-meson and the direct photon productions in the $pp$
collision\,\cite{KY11,KY11direct}. 
We will derive the corresponding single-spin dependent cross sections 
by applying the formalism 
developed for the semi-inclusive deep inelastic scattering\,\cite{BKTY10}.
We will also present a model estimate for the triple-gluon correlation functions
by comparing our result with the RHIC preliminary data for the $D$-meson
production\,\cite{Liu}.  Finally we perform numerical calculation of
the asymmetry for the direct
photon production by using the models obtained from $p^\uparrow p\to DX$
to see its impact 
on the SSA for this process.

\section{2. triple-gluon correlation functions}

Triple-gluon correlation functions for the transversely 
polarized nucleon are defined as the color-singlet nucleon matrix element
composed of the three gluon's field strength tensors $F^{\alpha\beta}$.
Corresponding to the two structure constants for the color SU(3) group, 
$d_{bca}$ and $f_{bca}$, 
one obtains two independent triple-gluon correlation functions $O(x_1,x_2)$ and $N(x_1,x_2)$ 
as\,\cite{BKTY10}
\small
\beq
&&\hspace{-0.8cm}O^{\alpha\beta\gamma}(x_1,x_2)
=-g(i)^3\int{d\lambda\over 2\pi}\int{d\mu\over 2\pi}e^{i\lambda x_1}
e^{i\mu(x_2-x_1)}\la pS|d_{bca}F_b^{\beta n}(0)F_c^{\gamma n}(\mu n)F_a^{\alpha n}(\lambda n)
|pS\ra \nonumber\\
&&=2iM_N\left[
O(x_1,x_2)g^{\alpha\beta}\epsilon^{\gamma pnS}
+O(x_2,x_2-x_1)g^{\beta\gamma}\epsilon^{\alpha pnS}
+O(x_1,x_1-x_2)g^{\gamma\alpha}\epsilon^{\beta pnS}\right]
\label{3gluonO},\\
&&\hspace{-0.8cm}N^{\alpha\beta\gamma}(x_1,x_2)
=-g(i)^3\int{d\lambda\over 2\pi}\int{d\mu\over 2\pi}e^{i\lambda x_1}
e^{i\mu(x_2-x_1)}\la pS|if_{bca}F_b^{\beta n}(0)F_c^{\gamma n}(\mu n)F_a^{\alpha n}(\lambda n)
|pS\ra \nonumber\\
&&=2iM_N\left[
N(x_1,x_2)g^{\alpha\beta}\epsilon^{\gamma pnS}
-N(x_2,x_2-x_1)g^{\beta\gamma}\epsilon^{\alpha pnS}
-N(x_1,x_1-x_2)g^{\gamma\alpha}\epsilon^{\beta pnS}\right],
\label{3gluonN}
\eeq
\normalsize
where $M_N$ is the nucleon mass, $S$ is the transverse-spin vector 
for the nucleon, 
$n$ is the light-like vector satisfying $p\cdot n=1$ and
we used the shorthand notation as $F^{\beta n}\equiv
F^{\beta\rho}n_{\rho}$ {\it etc}.  The gauge-link operators which
restore gauge invariance of the correlation functions
are suppressed in (\ref{3gluonO}) and (\ref{3gluonN}) for simplicity.  

\section{3. $D$-meson production in $pp$ collision}

Applying the formalism for
the contribution of the triple-gluon correlation functions to SSA developed in \cite{BKTY10}, 
the twist-3 cross section for
$p^{\uparrow}(p,S_\perp) + p(p') \to D(P_h)  + X$
(center-of-mass energy $\sqrt{S}$) can be obtained in the following form\,\cite{KY11}: 
\footnotesize
\beq
&&\hspace{-0.9cm}
P_h^0\frac{d\Delta\sigma}{d^3P_h}=\frac{\alpha_s^2M_N\pi}{S}\epsilon^{P_h p n S_{\perp}}
\sum_{f=c\bar{c}}\int\frac{dx'}{x'}G(x')\int\frac{dz}{z^2}D_f(z)\int\frac{dx}{x}\delta
 \left(\tilde{s}+\tilde{t}+\tilde{u}\right){1\over z\tilde{u}}
 \nonumber\\
&&\hspace{-0.9cm}
\times\biggl[\delta_f\left\{
\left(\frac{d}{dx}O(x,x)-\frac{2O(x,x)}{x}\right)\hat{\sigma}^{O1}
+\left(\frac{d}{dx}O(x,0)-\frac{2O(x,0)}{x}\right)\hat{\sigma}^{O2}
+\frac{O(x,x)}{x}\hat{\sigma}^{O3}
+\frac{O(x,0)}{x}\hat{\sigma}^{O4}
\right\} \nonumber\\
&&\hspace{-0.9cm}
+\left\{
\left(\frac{d}{dx}N(x,x)-\frac{2N(x,x)}{x}\right)\hat{\sigma}^{N1}
+\left(\frac{d}{dx}N(x,0)-\frac{2N(x,0)}{x}\right)\hat{\sigma}^{N2}
+\frac{N(x,x)}{x}\hat{\sigma}^{N3}
+\frac{N(x,0)}{x}\hat{\sigma}^{N4}
\right\}
\biggr],
\label{twist3final}
\eeq
\normalsize
where 
$\delta_c=1$ and 
$\delta_{\bar{c}}=-1$, $D_f(z)$ represents the 
$c\to D$ or $\bar{c}\to\bar{D}$ fragmentation functions, $G(x')$ is the
unpolarized gluon density, $p_c$ is the four-momentum of the $c$ (or
$\bar{c}$) quark (mass $m_c$) fragmenting into the
final $D$ (or $\bar{D}$) meson and 
$\tilde{s}$, $\tilde{t}$, $\tilde{u}$ are defined as $
\tilde{s}=(xp+x'p')^2,~\tilde{t}=(xp-p_c)^2-m_c^2,~\tilde{u}=(x'p'-p_c)^2-m_c^2.$
The hard cross sections $\hat{\sigma}^{O1,O2,O3,O4}$ and
$\hat{\sigma}^{N1,N2,N3,N4}$ are listed in \cite{KY11}.  
The cross section (\ref{twist3final}) receives contributions from
$O(x,x)$, $O(x,0)$, $N(x,x)$ and $N(x,0)$ separately, which differs from the previous
result\,\cite{KQVY08}. 

We perform numerical estimate for $A_N$ based on (\ref{twist3final}).
Since $|\hat{\sigma}^{O3,O4,N3,N4}|\ll |\hat{\sigma}^{O1,O2,N1,N2}|$
and $\hat{\sigma}^{O1} \simeq \hat{\sigma}^{O2} \sim \hat{\sigma}^{N1} \simeq -\hat{\sigma}^{N2}$, 
we assume the relation for the four functions as
$O(x,x)=O(x,0)=N(x,x)=-N(x,0)$ for simplicity. 
For the functional form of each functions, we employ the following two models:  
\beq
&&{\rm Model\ 1}:\qquad O(x,x)=K_G\,x\,G(x),
\label{model1}\\
&&{\rm Model\ 2}:\qquad O(x,x)=K_G'\,\sqrt{x}\,G(x), 
\label{model2}
\eeq
where $K_G$ and $K_G'$ are the constants to be determined so that the
calculated asymmetry is consistent with the RHIC data\,\cite{Liu}.

For the numerical calculation, we use GJR08 \cite{GJR08}
for $G(x)$ and KKKS08 \cite{KKKS08} for $D_f(z)$.  We calculate $A_N$
for the $D$ and
$\bar{D}$ mesons at the RHIC energy at $\sqrt{S}=200$
GeV and the transverse momentum of the $D$-meson $P_T=2$ GeV.
We set the scale of all the distribution and fragmentation functions at
$\mu=\sqrt{P_T^2+m_c^2}$ with the charm quark mass $m_c=1.3$ GeV.

\begin{figure}[h]

\scalebox{0.35}{\includegraphics{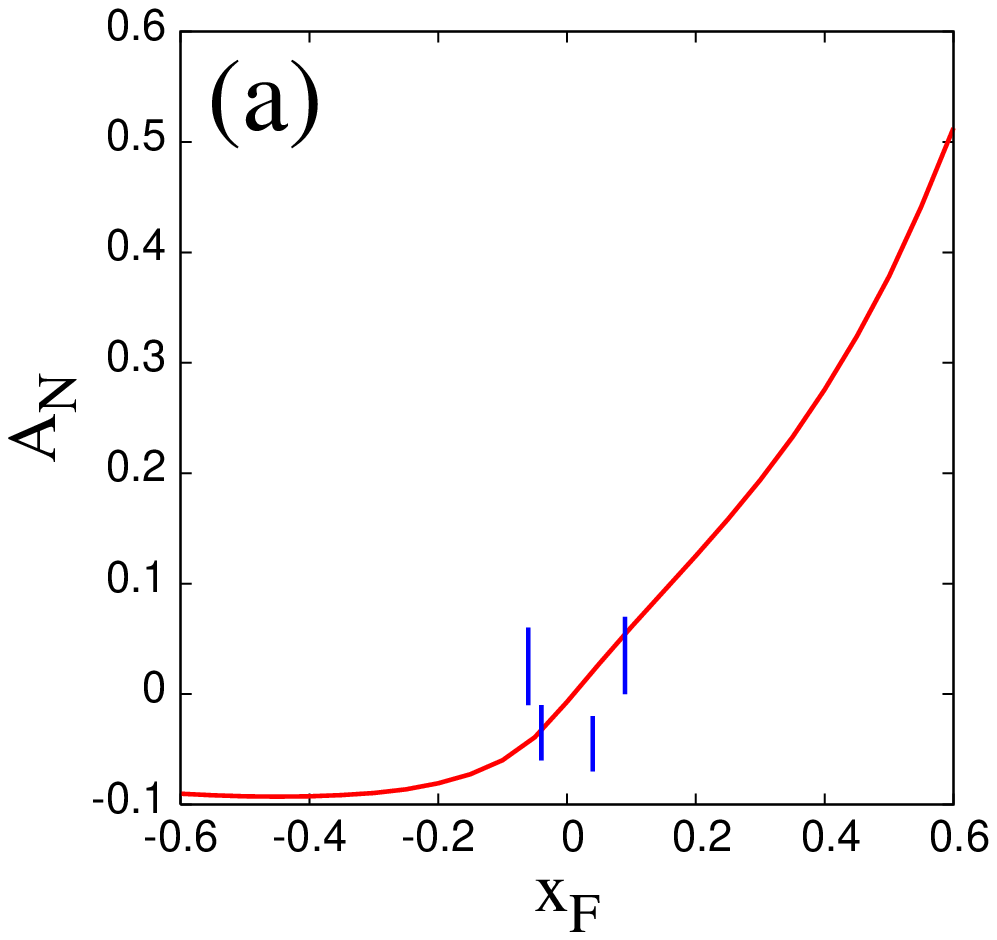}}
\scalebox{0.35}{\includegraphics{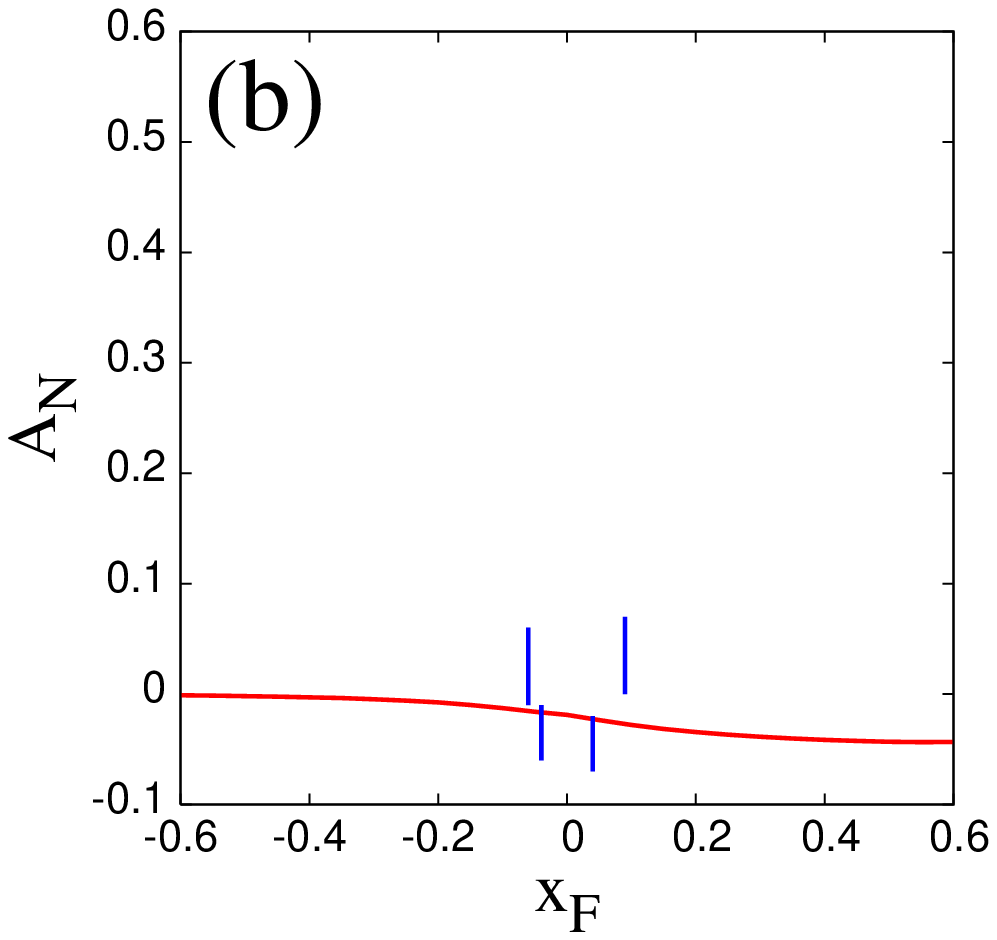}}

\scalebox{0.35}{\includegraphics{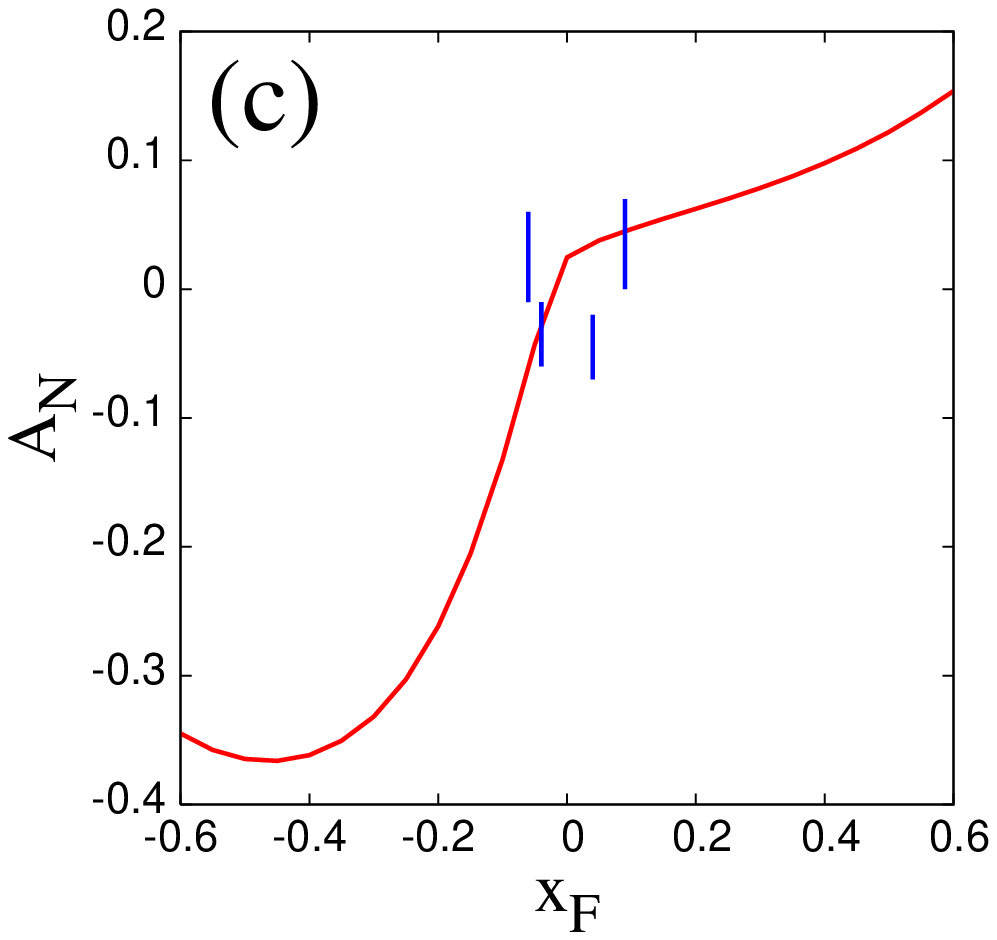}}
\scalebox{0.35}{\includegraphics{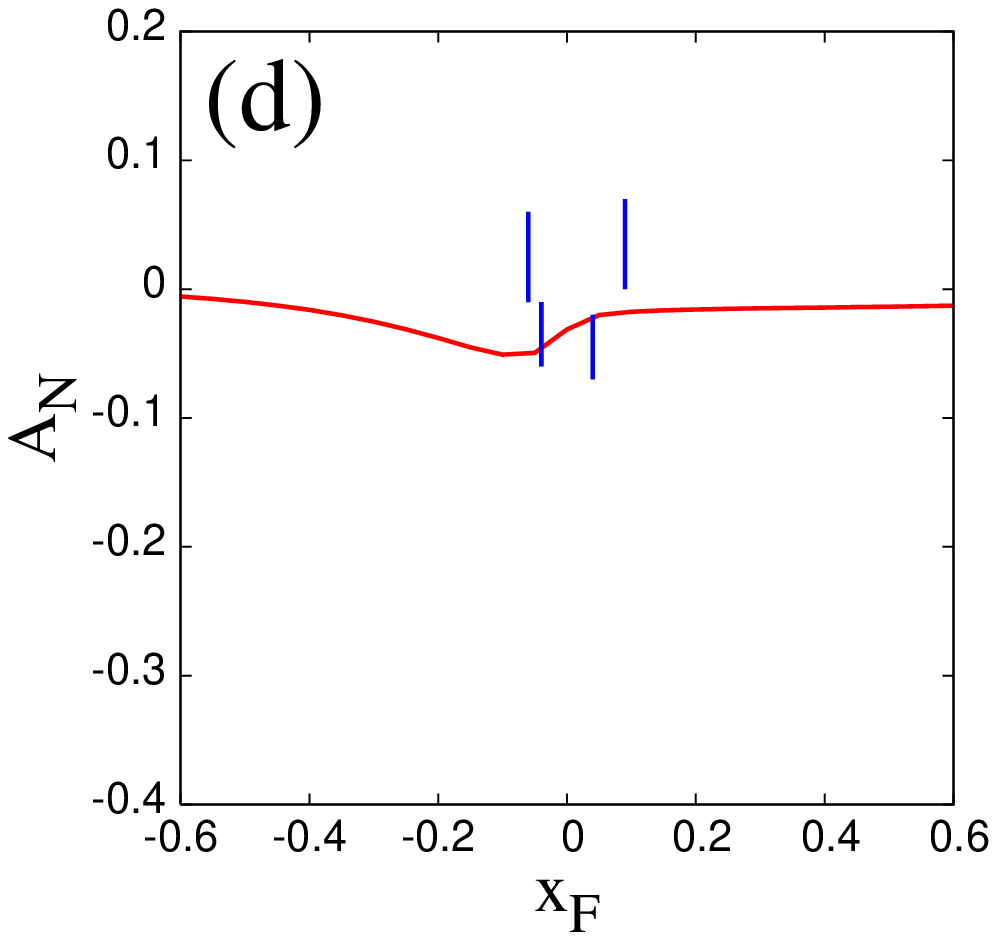}} 

 \caption{Results of $A_N^D$ for $D^0$ (a) and $\bar{D^0}$ (b)
 for Model 1 in (\ref{model1}) with $K_G=0.002$, and $A_N^D$ for
 $D^0$ (c) and $\bar{D^0}$ (d) for Model 2 in
 (\ref{model2}) with $K_G'=0.0005$.  Short bars denote the RHIC preliminary data
 taken from \cite{Liu}.}
 \label{fig:one}

\end{figure}

Fig. 1 shows the result of $A_N$ for the $D^0$ and $\bar{D}^0$ 
mesons together with the preliminary data\,\cite{Liu} denoted by the
short bars.  
The sign of the contribution from $\{O(x,x),O(x,0)\}$ changes
between $D^0$ and $\bar{D}^0$ as shown in (\ref{twist3final}), which causes the
large difference between $A_N$ for the $D^0$ and $\bar{D}^0$.
If one reverses the relative sign between $O$ and $N$,
the result for the $D^0$ and $\bar{D}^0$ mesons will be interchanged.  
The values $K_G=0.002$ and $K'_G=0.0005$ have been determined so that
$A_N$ does not overshoot the RHIC data.  
By comparing the results for the models 1 and 2 in Fig. 1, 
one sees that the behavior of $A_N$ at $x_F<0$ depends strongly on
the small-$x$ behavior of the triple-gluon correlation functions.  
Therefore $A_N$ at $x_F<0$ is useful to get constraint on the
small-$x$ behavior of the three-gluon correlation functions.  

\section{4. direct photon production in $pp$ collision}

Applying the same formalism, 
the twist-3 cross section for the direct photon production, 
$p^{\uparrow}(p,S_\perp) + p(p') \to \gamma(q)  + X$, 
induced by the triple-gluon correlation functions
can be obtained as\,\cite{KY11direct}
\small
\beq
 E_{\gamma}\frac{d\sigma}{d^3q}&=&\frac{4\alpha_{em}\alpha_sM_N\pi}{S}\sum_{a}\int\frac{dx'}{x'}f_a(x')\int\frac{dx}{x}\delta
 (\hat{s}+\hat{t}+\hat{u})\epsilon^{q p n S_{\perp}}{1\over \hat{u}}
 \nonumber\\
&& \times\biggl[\delta_a
\Bigl(\frac{d}{dx}O(x,x)-\frac{2O(x,x)}{x}
+\frac{d}{dx}O(x,0)-\frac{2O(x,0)}{x}\Bigr) \nonumber\\
&&
-\frac{d}{dx}N(x,x)+\frac{2N(x,x)}{x}
+\frac{d}{dx}N(x,0)-\frac{2N(x,0)}{x}\biggr]
\left({1\over N}{\hat{s}^2+\hat{u}^2\over \hat{s}\hat{u}}\right), 
\label{twist3DP}
\eeq
\normalsize
where $f_a(x')$ is the twist-2 unpolarized quark density,
$\delta_a=1(-1)$ for quark (antiquark) and 
$\hat{s}$, $\hat{t}$, $\hat{u}$ are defined as $\hat{s}=(xp+x'p')^2,~\hat{t}=(xp-q)^2,~\hat{u}=(x'p'-q)^2$.  
As shown in (\ref{twist3DP}), the combinations $O(x,x)+O(x,0)$ and
$N(x,x)-N(x,0)$ appear in the
cross section accompanying the common partonic hard cross section 
which is the same as the twist-2 hard cross section
for the $qg\to q\gamma$ scattering.  
This result differs from the previous
study in \cite{Ji92}. 

We performed a numerical calculation for $A_N^\gamma$ for the following two cases: 
Case 1; $O(x,x)=O(x,0)=N(x,x)=-N(x,0)$ and 
Case 2; $O(x,x)=O(x,0)=-N(x,x)=N(x,0)$. 
We use GJR08 \cite{GJR08} for 
$f_q(x')$ and the models (\ref{model1}) and (\ref{model2}) with $K_G=0.002$ and
$K_G'=0.0005$ which are consistent with RHIC $A_N^D$ data.
We calculate $A_N^\gamma$ at
the RHIC energy at $\sqrt{S}=200$ GeV and the
transverse momentum of the photon $q_T=2$ GeV, 
setting the scale of all the distribution at $\mu=q_T$.

\begin{figure}[h]

\scalebox{0.35}{\includegraphics{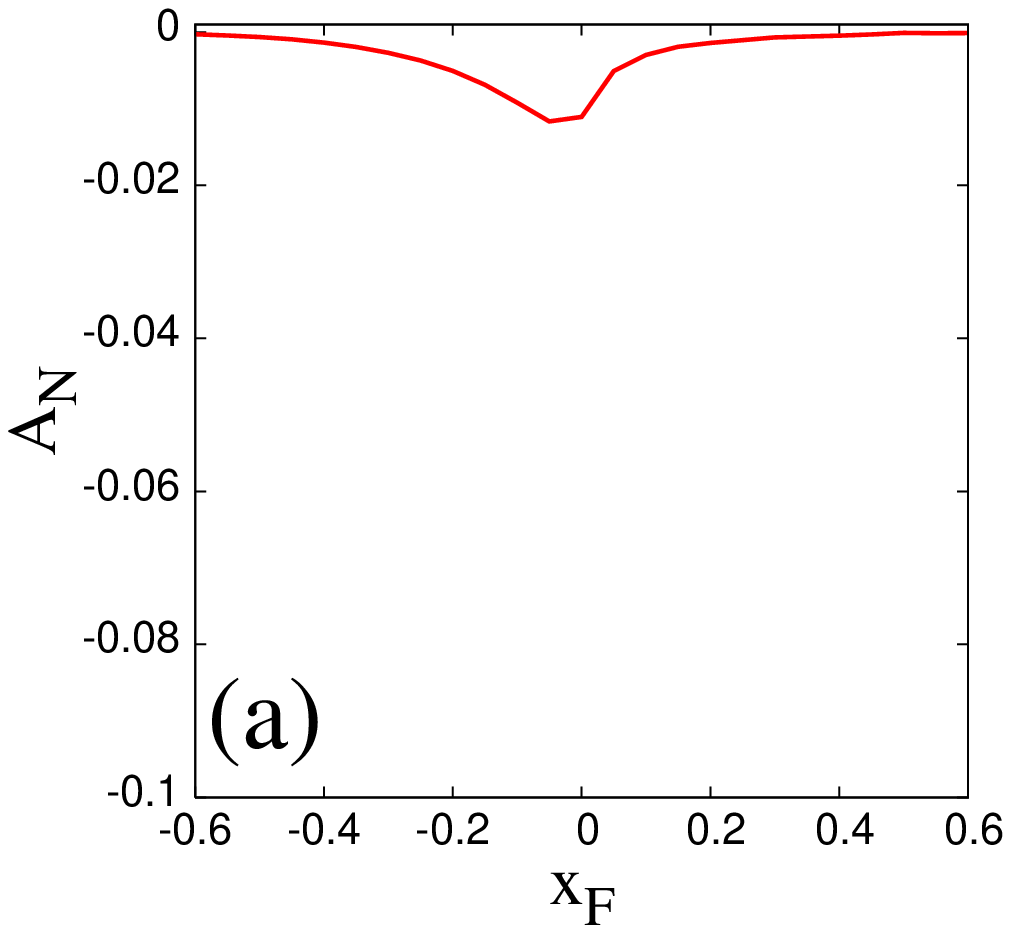}}
\scalebox{0.35}{\includegraphics{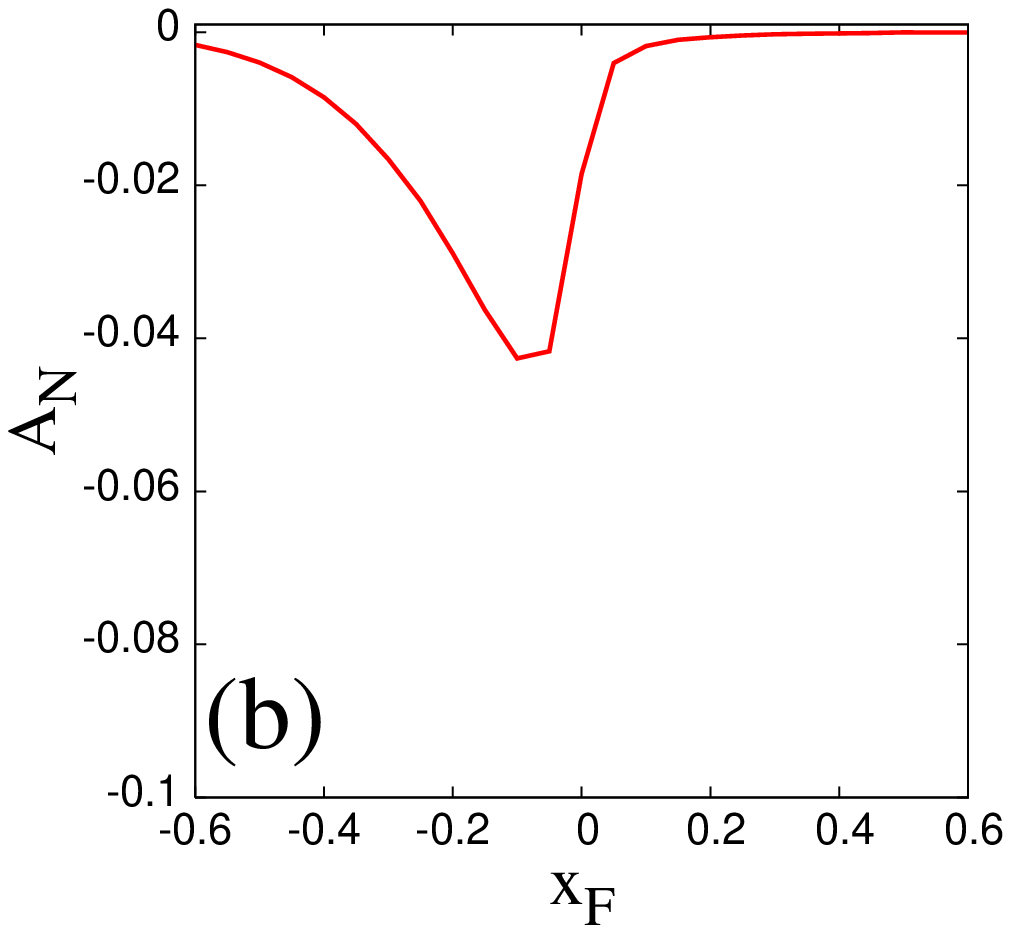}}

\scalebox{0.35}{\includegraphics{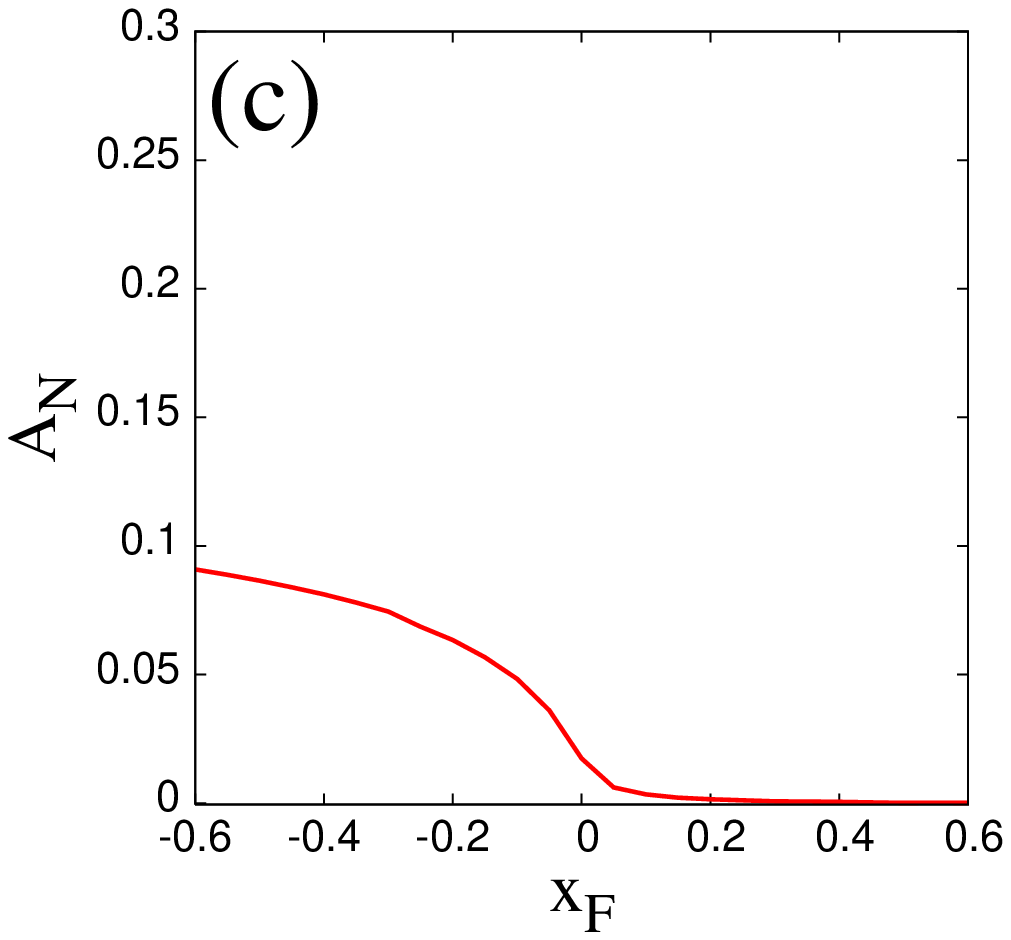}}
\scalebox{0.35}{\includegraphics{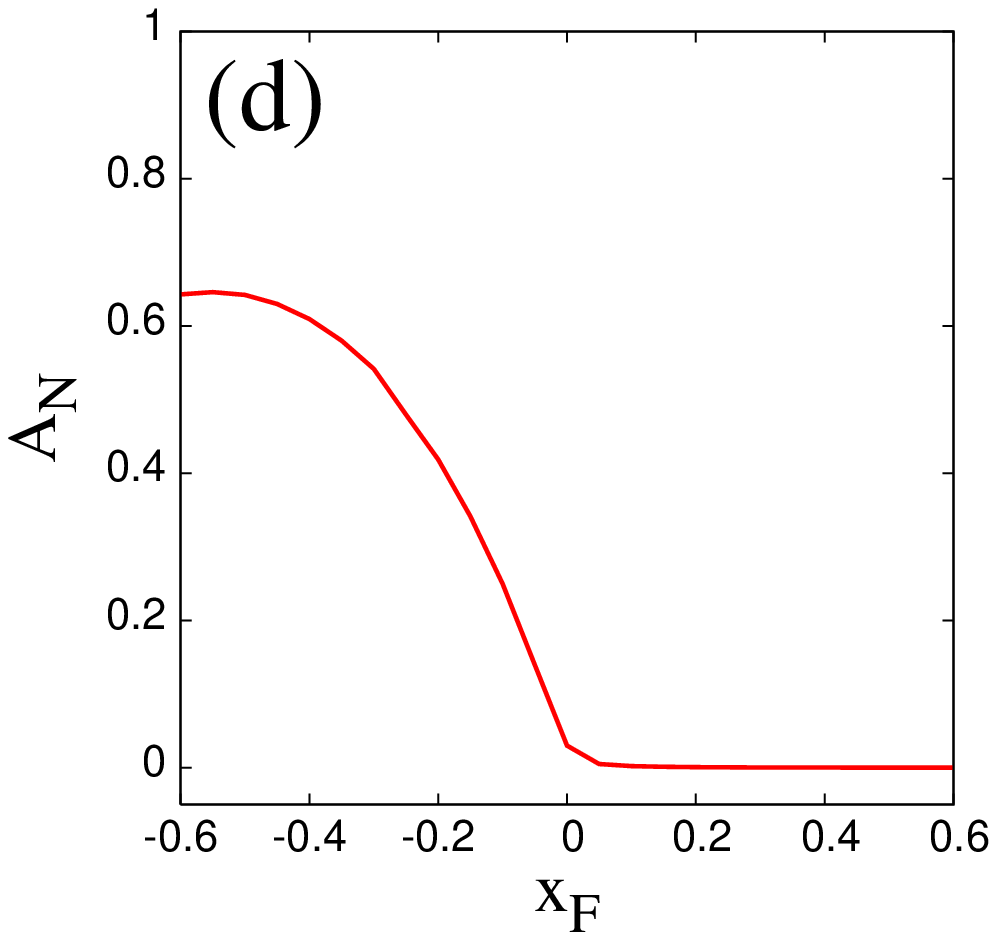}}

 \caption{(a) $A_N$ for Case 1 with Model 1. (b) $A_N$ for Case 1 with Model 2. 
(c) $A_N$ for Case 2 with Model 1. (d) $A_N$ for Case 2 with Model 2. }
 \label{fig:two}

\end{figure}

Fig. 2 shows the result for $A_N^\gamma$ for each case. One can see
$A_N$ at $x_F> 0$ become almost zero regardless of the magnitude of
the triple-gluon correlation functions, 
while $A_N$
at $x_F<0$ depends strongly on the small-$x$ behavior of the
triple-gluon correlation functions as in the case of $p^{\uparrow}p\to DX$.
At negative $x_F$, large-$x'$ region of the unpolarized quark distributions
and the small-$x$ region of the triple-gluon distributions are relevant.  
For the above case 1, only antiquarks in the unpolarized nucleon are active
and thus lead to small $A_N^\gamma$ as shown in Figs. 2(a) and (b).  
On the other hand, 
for the case 2, quarks in the unpolarized nucleon are active and thus
lead to large $A_N^\gamma$ as shown in  Figs. 2(c) and (d).  
Therefore $A_N^\gamma$ at $x_F<0$ for the direct photon
production could provides us with an important information on the relative sign between $O$
and $N$.




\begin{theacknowledgments}
This work is supported by the Grand-in-Aid for Scientific Research
(No. 23540292 and No. 22.6032) from the Japan Society for the Promotion of Science.
\end{theacknowledgments}



\bibliographystyle{aipproc}   

\bibliography{sample}

\IfFileExists{\jobname.bbl}{}
 {\typeout{}
  \typeout{******************************************}
  \typeout{** Please run "bibtex \jobname" to optain}
  \typeout{** the bibliography and then re-run LaTeX}
  \typeout{** twice to fix the references!}
  \typeout{******************************************}
  \typeout{}
 }

\end{document}